# Digital Divide in Disasters: Investigating Spatial and Socioeconomic Disparities in Internet Service Disruptions During Extreme Weather Events

Yuvraj Gupta[1], Zhewei Liu[2], Ali Mostafavi[2]

[1] Department of Civil Engineering, Indian Institute of Technology Gandhinagar
[2] UrbanResilience.AI Lab, Zachry Department of Civil and Environmental Engineering, Texas A&M University, College Station, TX, 77843

**Abstract**
The resilience of internet service is crucial for ensuring consistent communication, situational awareness, facilitating emergency response in our digitally-dependent society. However, due to empirical data constraints, there has been limited research on internet service disruptions during extreme weather events. To bridge this gap, this study utilizes observational datasets on internet performance to quantitatively assess the extent of internet disruption during two recent extreme weather events. Taking Harris County in the United States as the study region, we jointly analyzed the hazard severity and the associated internet disruptions in the context of two extreme weather events. The results show that the hazard events significantly impacted regional internet connectivity. There exists a pronounced temporal synchronicity between the magnitude of disruption and hazard severity: as the severity of hazards intensifies, internet disruptions correspondingly escalate, and eventually return to baseline levels post-event. The spatial analyses show that internet service disruptions can happen even in areas that are not directly impacted by hazards, demonstrating that the repercussions of hazards extend beyond the immediate area of impact. This interplay of temporal synchronization and spatial variance underscores the complex relationships between hazard severity and Internet disruption. Furthermore, the socio-demographic analysis suggests that vulnerable communities, already grappling with myriad challenges, face exacerbated service disruptions during these hazard events, emphasizing the need for prioritized disaster mitigation strategies and interventions for improving the resilience of internet services. To the best of our knowledge, this research is among the first studies to examine the Internet disruptions during hazardous events using a quantitative observational dataset. The insights obtained hold significant implications for city administrators, guiding them towards more resilient and equitable infrastructure planning.

# 1. Introduction

Information and communication technology (ICT) and its underlying service have become vital components of modern society, supporting daily digital activities from commerce and communication to critical services and emergency management. The robustness of this infrastructure, however, becomes a significant concern when faced with natural hazards (Coleman, 2020; Esmalian, 2021; Yin, 2023a). As indicated in multiple studies, extreme weather and hazard events can lead to cascading impacts across critical infrastructure, the repercussions of which ripple across sectors, affecting access to these critical services and altering perceptions of recovery. (Han et al., 2009; Hasan and Foliente, 2015; FEMA, 2017; Zimmerman et al., 2017; Mitsova et al., 2020).

Historically, a multitude of disasters have exposed the vulnerabilities in our critical infrastructure systems, prompting significant research into their resilience. For instance, Hurricane Irma disrupted electricity for millions across Florida, from the Keys to the Panhandle, demonstrating how such a widespread outage can significantly strain recovery efforts (Davis et al., 2019). Meanwhile, Hurricane Michael wreaked havoc in the Mexico Beach area of Florida, damaging more than 700 structures and leaving a large number of people without electricity for nearly a month. The aftermath of Hurricane Maria in Puerto Rico was particularly devastating. The hurricane severely impacted an already fragile infrastructure system, resulting in widespread loss of services. Remarkably, almost a month post-disaster, 90% of Puerto Rico's households remained without power, and many lacked water or cell phone service, precipitating a humanitarian crisis that lingered for months (Roman 2018; Zorrilla 2018). Beyond these hurricanes, other disasters also highlighted infrastructure vulnerabilities. In the 1995 Kobe earthquake, disruptions in firefighting capabilities, primarily due to a lack of pressurized water, led to the rapid spread of urban fires causing extensive human and property losses. Furthermore, disruptions to road networks can inhibit emergency responses, restricting the movement of critical emergency services, including firefighters, ambulances, and utility repair crews (Kirsch et al., 2010; Yin,. 2023b). These examples underscore the necessity of building resilient infrastructures, emphasizing the lessons learned from hazard events (Coleman et al., 2020; Fan and Mostafavi, 2019; Zhang et al., 2020).

Despite the extensive studies and insights on infrastructure resilience during disasters, a significant research gap persists regarding the understudied resilience of internet service. Previous studies have highlighted significant disparities in internet access across various communities, with socially vulnerable groups—including racial minorities, low-income families, and rural residents—facing pronounced challenges (Graves et al., 2021; Chiou and Tucker, 2020; Ho, 2023). These challenges are limited not only to inadequate device ownership but extend to spotty cellular coverage and the unaffordability of data plans. These disparities in digital access have exacerbated the impaired living conditions of these vulnerable communities, causing reduced work efficiency, reduced access to news and commerce, and limited access to education during the pandemic. Such challenges further intensify existing societal inequalities (Sen and Tucker 2020; Singh et al., 2020; Liu, 2023; Huang, 2022). While there has been considerable research on internet access disparities under normal circumstances, there has been limited investigation into these disparities during times of disaster. Emphasizing the vital role of communications during disasters, studies indicate that communication systems play pivotal roles in information sharing and protective actions during disaster response and recovery (Zhang et al., 2019; Fan et al., 2021; Yin, et al. 2012). Given the intricacies of measuring disruption, especially for communication systems, there is a dearth of data-driven approaches to assessing the resilience of internet service ( Mattson and Jenelius, 2015). With the ever-increasing reliance on internet technology, there is a pressing need to delve deeper into its vulnerabilities and to establish measures that maintain its resilience during catastrophic events, ensuring continued access and functionality even in the face of adversity.

Consequently, addressing the above-mentioned research gaps, this study utilized observational data for internet performance provided by internet service provider to assess the extent of internet disruption during major weather events. Specifically, we take Harris County, in which Houston is located, as the study region and evaluate the impacts of two recent major hazard events, Hurricane Harvey (2017) and Winter Storm Uri (2021), on regional internet service. By analyzing how the internet is disrupted during these hazards, we aim to quantify the relationship between the severity of a hazard and the magnitude of internet disruption and to shed light on how different communities, especially socially vulnerable populations, are affected by these disruptions and ICT disconnectivity in the context of extreme events.

Particularly, we focus on the following research questions:
- **RQ 1:** To what extent is internet disruption associated with hazard intensity? Do regions that experience greater hazard severity also face more pronounced internet disruptions, leading to increased connectivity isolation?
- **RQ 2:** To what extent do areas with vulnerable communities have disproportional internet disruption? Will internet disruption exacerbate the precarious living conditions of vulnerable communities and hence cause new injustice?

By answering the above questions, this study investigates the effects of extreme hazards on internet connectivity and the potential emergence of social inequities that may disproportionately impact vulnerable communities. The findings of this study offer contributions across several dimensions: (1) To the best of our knowledge, this is the first study to examine internet disruptions during hazardous events using a comprehensive, quantitative observational dataset. The data provided by the internet service provider offers a distinctive view that enables us to model hazard-induced Internet disruption from a quantitative perspective, which has not been addressed by previous studies; (2) Our findings provide novel empirical evidence on the extent to which different communities (especially socially vulnerable communities) are affected by internet disruption during extreme hazard events and highlight the potential risks of emerging social injustices stemming from these disruptions; (3) The findings of study show the temporal synchronicity and spatial correlations between hazard severity and Internet disruption, which provide urban planners, city managers, and infrastructure operators with novel insights on crafting timely and effective strategies to alleviate environmental hazards and inequalities, and ultimately contribute to broader urban resilience and sustainability.

## 2. Dataset and Methodology

### 2.1 Dataset

This study aims to examine the relationship between internet disruption and hazard severity during extreme weather events, as well as the effect of internet service disruptions on social disparities. This study focuses on the city of Houston, the fourth most populous city in the US, which is exposed to frequent and severe weather disturbances due to its proximity to the Gulf of Mexico. Specifically, we focus on two recent major hazard events in the regions: Winter Storm Uri and Hurricane Harvey. Winter Storm Uri hit Houston in February 2021, crippling power grids and leaving millions without electricity in freezing conditions. Hurricane Harvey, which hit in 2017, brought with it torrential rain, causing unprecedented flooding in Houston. The storm's prolonged stay and intense rainfall turned streets into rivers and inundated vast portions of the city.

To quantify the internet performance and hazard severity, the following four datasets were used for analysis, as shown in Table 1:

- *Internet Connectivity:* The dataset contains the metrics for measuring the internet access performance. Our dataset is provided by Ookla, which provides free analysis for internet performance evaluation (Ookla, 2023). The provided dataset covers information regarding cellular internet speeds, detailing metrics such as a mobile device's upload/download rates, latency, and geographically pertinent data including the locations of both the device and the server. In this research, the upload and download speed at the client location is used as a metric for internet connection.
- *Power outage:* The magnitude of power outages during these events is extrapolated from telemetry-based population activity proffered by Mapbox. This service collects user cell phone locations from applications harnessing the Mapbox software development kit (SDK), subsequently aggregating, standardizing, and anonymizing this geographical data to estimate population activity. Numerous studies have employed Mapbox's population activity data to illuminate population dynamics during disaster events (Yuan et al., 2022; Gao et al., 2021; Farahmand et al., 2022; Lee et al, 2022). In this study, data was collected for February 2021, and the power outage used as a measurement of the hazard severity during Winter Storm Uri.
- *Flood extent:* To evaluate the spatial variation of lifestyle impacts from flooding status, we used flooding data from the estimated flood depths on August 29, 2017 (FEMA, 2018). The data had a gridded horizontal resolution of three meters, which was processed appropriately for the census block group (CBG)-scale analysis. The flood extent is used as a measurement of the hazard severity for Hurricane Harvey.
- *Sociodemographic data:* To assess the impact of internet disruption on different communities, Zip-code-level statistics data (e.g., total population, below-poverty population, minority population, etc.) is provided by the United States Census Bureau (USCB, 2023)

**Table 1**. Description for the used dataset

| Data | Data Source | Description |
|---|---|---|
| Internet connectivity | Ookla | Internet upload and download speed at specific locations |
| Hazard severity measured by power outage | MapBox | The extent of power outage in the areas is inferred from the telemetry-based population activity |
| Hazard severity measured by flood extent | Federal Emergency Management Agency (FEMA) | The estimated flood extent for Hurricane Harvey |
| Sociodemographic | United States Census Bureau (USCB) | Statistics data for each census tract, including total population, income, minority population, etc. |

## 2.2 Methodology

### 2.2.1. Quantification of the degree of internet disruption and hazard severity

The quantifications of internet disruption and hazard intensity were needed to model their mutual relationships.

- For internet disruption, this study considered parameters linked to internet speed, such as upload and download speed. A decline in internet speed serves as an indicator of disruption. Specifically, to account for daily variations in speed, we calculated the average internet speed prior to the hazard event, treating it as a baseline. If, during a hazard event, either the upload or download speed dropped by more than 10% compared to this baseline, the internet in the affected region was deemed to be disrupted.
- For hazard severity, due to data availability, we adopted different metrics for each event. For Hurricane Harvey, we used the flood extent within each Zip code to gauge hazard intensity. For the winter storm Uri, the metric employed was the power-related parameter activity density (DA) as detailed in Section 2.1. Similar to the internet disruption criteria, if the DA value during the disaster decreases by 10% or more from its pre-disaster average, we consider the respective Zip code to have faced disruption.

All these metrics are assessed at the Zip code level.

### 2.2.2. Correlation modeling between internet disruption and hazard intensity

To understand the relationship between internet disruption and hazard intensity, we employed correlation analysis, with specific emphasis on the Pearson correlation coefficient (often represented by Pearson's R). This coefficient evaluates the linear association between two variables:

$$R = \frac{\sum (X_i - \bar{X})(Y_i - \bar{Y})}{\sqrt{\sum (X_i - \bar{X})^2 \sum (Y_i - \bar{Y})^2}}$$

where $X_i$ and $Y_i$ are the individual data points for the two datasets, $\bar{X}$ and $\bar{Y}$ are the means of the two datasets, the numerator is the sum of the products of the differences between each data point and its respective mean for both datasets. The denominator is the product of the square roots of the sums of the squared differences between each data point and its mean for dataset. The correlation coefficient *R* produced by this formula lies between -1 and 1; r=1 indicates a perfect positive linear relationship; r= -1 indicates a perfect negative linear relationship; r=0 indicates no linear relationship. By applying Pearson's R to our data on internet disruption and hazard intensity, we can infer the strength and direction of their linear relationship.

## 3. Results

### 3.1 The association and synchronicity between internet disruption and hazard intensity

We evaluated the relationships between hazard intensity and internet disruption. For Winter Storm Uri, the extent of power outages served as a metric for hazard severity. For Hurricane Harvey, we utilized flood extent as the proxy for hazard severity. The time series analysis shows that hazard severity exhibits a strong correlation with the disruption of internet connectivity. As illustrated in Figure 1, the temporal patterns between power outages and internet connection disruptions are significantly consistent. Specifically, the onset of power disruptions in Winter Storm Uri started in February 13, which coincides with the impact of the storm across the affected regions. The lowest point of these outages was observed on February 15, after which there was a gradual restoration of power. The internet connection's trajectory was similar to this, with a slight temporal shift. Internet connectivity began waning on February 14, a day after the initiation of managed power outages. This decline continued until February 16, at which point a revival started. Interestingly, this recovery in internet connectivity lagged the power restoration pattern by a day. This one-day delay between the two variables suggests a potential dependency of internet connectivity on hazard intensity or other intermediary factors that could be influenced by power availability. Even after primary issues (like power) are addressed, it may take additional time for secondary systems (like the internet) to regain full functionality.

The time series correlation analysis further quantifies the relationship between power outages and internet connectivity. Specifically, the correlation coefficients are 0.763 for download speeds and 0.774 for upload speeds. Both values are statistically significant at the 0.01 level. This underscores that as power outages became more pronounced, internet speeds, both download and upload, were adversely impacted. The high correlation values, combined with the temporal patterns observed in Figure 1, clearly demonstrate that internet connectivity is inherently vulnerable to power disruptions, especially during events like Winter Storm Uri.

The intertwining synchronicity of power outages and internet disruptions during large-scale extreme weather events reveals the interdependencies between power and communication infrastructure and also underscores a pressing concern: at times when communities most require essential services for situational awareness and information dissemination, they face the most significant disruptions. This poses a substantial challenge to both individuals and emergency services, exacerbating an already challenging situation. The onset of power outages, starting from February 13, aligns closely with the dwindling of internet connectivity, which began a day later. This interdependency indicates that when large-scale events strike, crucial infrastructure systems like power and the internet are both vulnerable, responding in near-tandem to the external threat. This cascading effect has profound implications for situational awareness: lack of access to critical information during the most crucial times can lead to panic, misinformation, and an increased reliance on already stretched emergency services.

In addition, the slight delay observed in the restoration of internet connectivity, even after power was reinstated post-Uri, hints that restoring primary systems, such as power, does not immediately guarantee the revival of secondary systems. This lag further exacerbates the challenges faced by affected communities, prolonging their state of vulnerability and isolation.

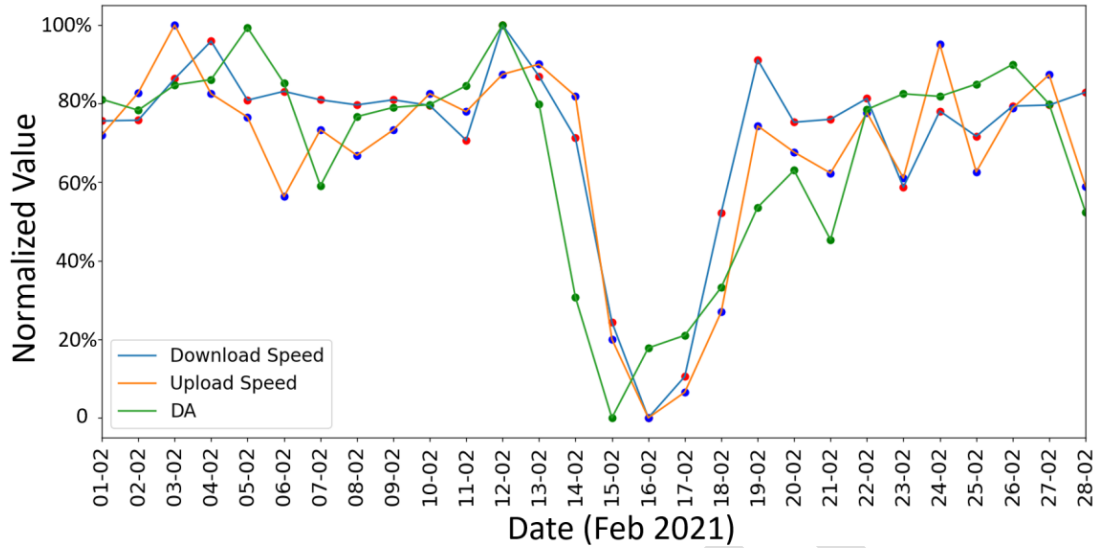

**Figure 1**. Time series of average internet speed and power supply (DA) in Houston during the Winter Storm Uri series analysis. The correlation coefficient between DA and download speed is 0.763 and upload speed is 0.774 (both statistically significant at 0.01 level), demonstrating the synchronousness and interdependency between internet connectivity and power supply.

We then looked into the spatial patterns of internet disruptions and power outages. When assessing the data between February 15 and February 18, 2021 (the timeframe impacted by Winter Storm Uri), we calculated the average degree of both power outage and internet disruption. These averages were then plotted at the ZIP code-level (Figure 2). The visual interpretation suggests that the regions most significantly affected by internet disruptions and power outages do not exhibit considerable overlap. This observation is quantitatively reinforced by a Pearson's correlation coefficient of only 0.028, which is not statistically significant.

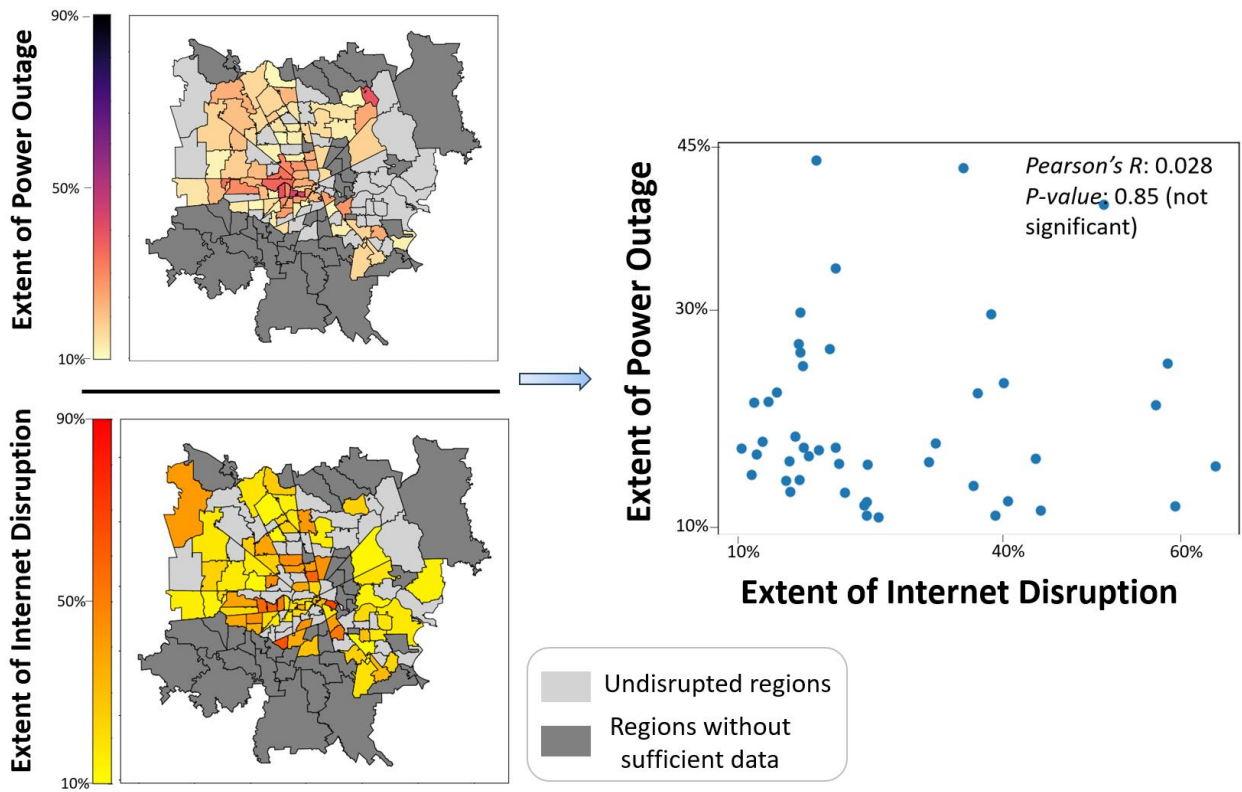

**Figure 2**. Spatial comparison between regions' extent of power outage and internet disruption during Winter Storm Uri.

A similar pattern was also observed for Hurricane Harvey. When comparing the average flood extent with internet disruption at the Zip code-level (Figure 3), distinct disparities arise. Specifically, the northeast regions of Houston experienced a more significant flood extent relative to other regions. Yet, the internet connections in these heavily flooded areas were less disrupted. The quantitative correlation analysis further supports this observation, with a Pearson's R value of 0.189, also proving to be not significant.

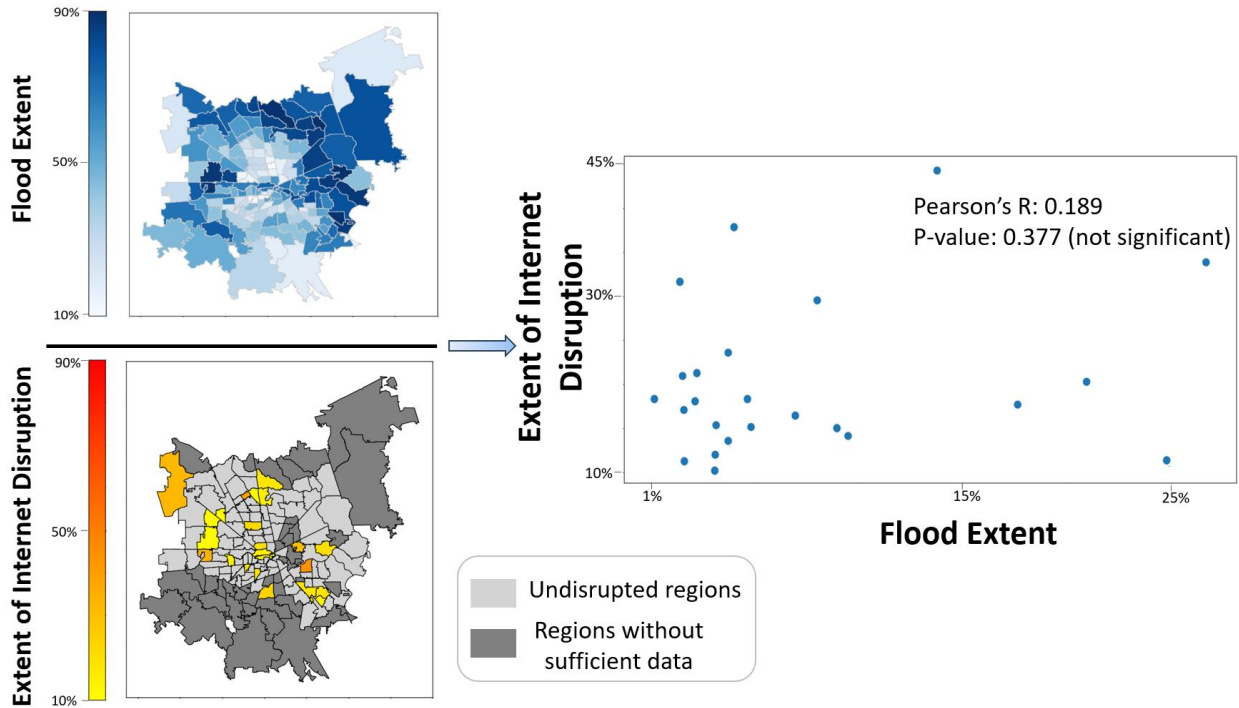

**Figure 3**. Spatial comparison between regions' extent of power outage and internet disruption during Hurricane Harvey

These spatial analyses underscore the phenomenon that regions with severe power outages or flooding may not always coincide with areas of significant internet disruption. This suggests the presence of other mitigating factors or service characteristics that safeguard internet connectivity despite surrounding hazards. For example, certain areas might be equipped with underground cabling systems that are less vulnerable to external hazards. Alternatively, some regions might have access to redundant internet service providers, ensuring that if one provider faces downtime, another can continue to offer service. Furthermore, localized service characteristics could play a pivotal role. Regions with advanced network services, such as modern data centers or state-of-the-art routing systems, might be better poised to withstand disruptions or to reroute traffic effectively during power outages or other hazards. There could also be localized backup power solutions, such as generators or battery reserves, specifically dedicated to maintaining internet service, ensuring a consistent online presence even when broader electrical grids are compromised.

The spatial disconnect between hazard intensity and internet disruption emphasizes the importance of not viewing internet connectivity in isolation. It is crucial to recognize that internet service disruptions can happen even in areas that are not directly impacted by hazards, which means that the repercussions of hazards extend beyond the immediate area of impact, affecting regions that might seem safe or untouched at first glance. This phenomenon underscores the broader implications of such hazards, as areas without direct hazard impacts can still experience disruptions in essential internet services. Disruptions of internet services in areas not impacted directly by the hazards can slow down the response and relief efforts to impacted areas due to the disrupted information dissemination and situational awareness.

While the temporal alignment between power outages and internet disruptions is evident, our spatial analyses offer a contrasting picture. The spatial patterns observed during both Winter Storm Uri and Hurricane Harvey exhibit a significant disconnect. Regions heavily affected by power outages or flooding did not necessarily overlap with areas facing the most internet disruptions. This spatial incongruence underscores the complexity of the infrastructure networks and their interdependencies. A potential explanation for this spatial inconsistency could be rooted in the spatial distribution of internet infrastructure (e.g., cell towers). Furthermore, the presence of redundant systems, be it in the form of multiple internet providers or backup power solutions, can greatly enhance a region's resilience, acting as a buffer against widespread disruptions. While the temporal consistency between hazard intensity and internet disruptions sheds light on the immediacy of the effects of large-scale events on infrastructural systems, the spatial inconsistency points towards a set of local factors, from infrastructure resilience to historical preparedness, that can mediate these effects. The findings call for a nuanced approach in disaster management, focusing not just on immediate responses but also on strengthening spatially diverse infrastructural systems, making them more resilient in the face of future hazards.

**3.2 Environmental Injustice: Social Demographic Analysis**

We further explored the matter of environmental injustice issues regarding internet disruption during hazard events. Particularly, the grouping of regions is made based on the residents' income level (Section 2.1): The calculation of the median of percentage of the population living below the poverty line was performed. Regions where this percentage exceeds the median are classified as low-income groups, whereas those with percentages below the median are designated as high-income groups. After grouping, the pre-hazard internet connectivity and the internet disruption of each income group during the hazard in each group is calculated.

As illustrated in Figure 4, during the events of Winter Storm Uri and Hurricane Harvey, low-income communities experienced reduced internet speeds and heightened internet disruptions (quantified by the percentage decrease in internet speed during these events) compared to high-income communities. Notably, this disparity was particularly pronounced during Winter Storm Uri. In this instance, the decline in internet speed for high-income communities was approximately 25% to 30%, while it exceeded 40% for low-income communities. This substantial difference underscores a marked disparity faced by low-income communities in terms of internet connectivity during such hazard events.

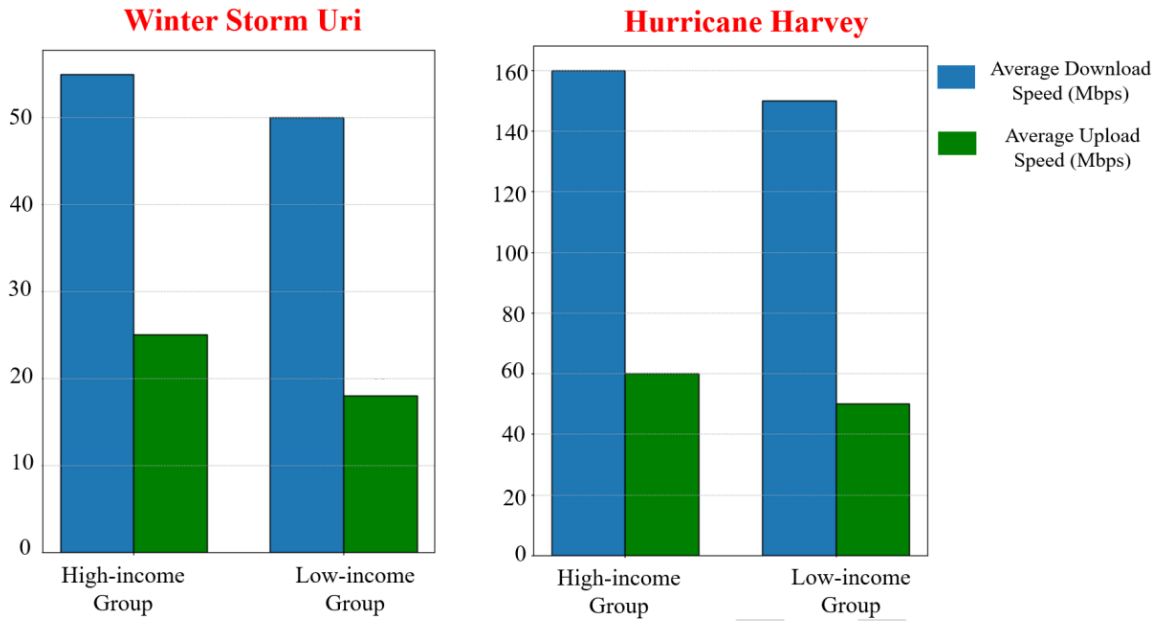
(a) Pre-hazard Internet Speed

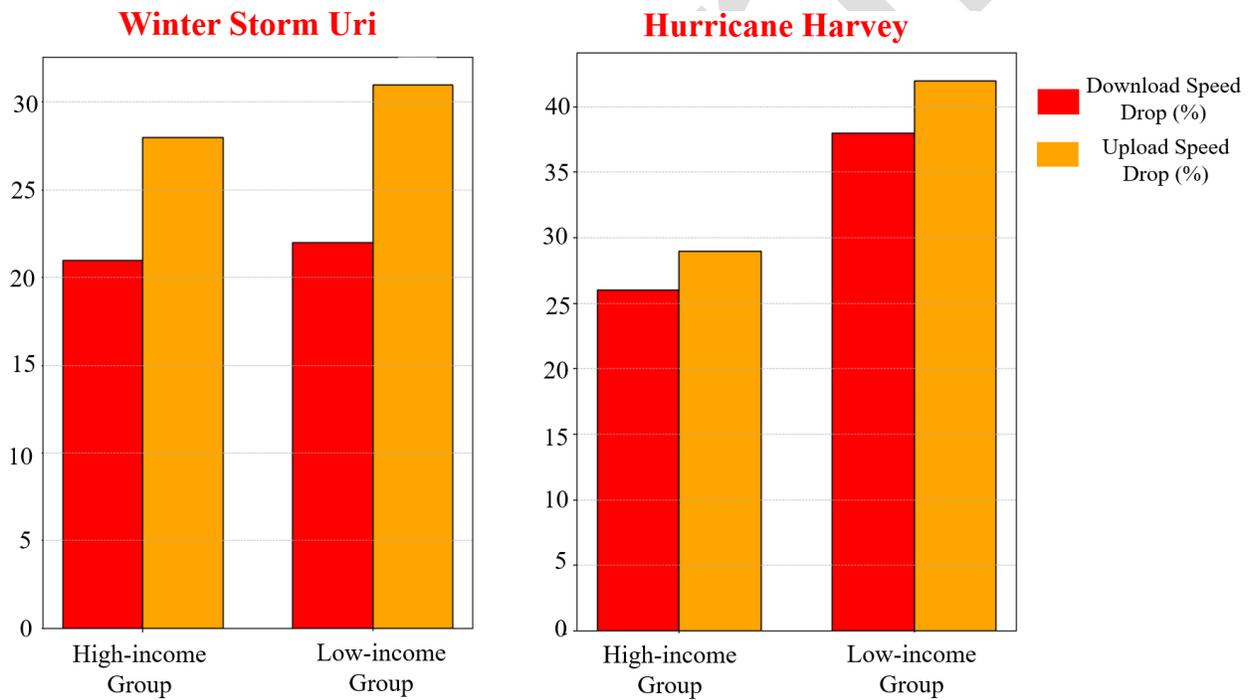
(b) Internet Speed Drop During Hazards

**Figure 4**. Pre-hazard internet connectivity and the internet disruption during the hazards.

The patterns uncovered from the data, specifically during Winter Storm Uri and Hurricane Harvey, bring forth an unsettling realization about the intricacies of internet disruptions in vulnerable communities. In times of crises, the internet serves as a lifeline for many, providing essential information on safety measures, weather updates, emergency services, and more. Situational awareness is crucial during such events, and the internet plays a pivotal role in disseminating real-time information. Vulnerable communities, often already grappling with socio-economic challenges, find themselves in an even more precarious situation when they cannot access critical information due to internet outages. Moreover, the internet is also a primary tool for information sharing. Communities use it to organize relief efforts, and individuals use it to seek help or provide updates about their safety. When internet services are disrupted, especially in areas with a high concentration of vulnerable populations, the effects can be detrimental to life and health.

The findings from Winter Storm Uri and Hurricane Harvey highlight that at times and in places where vulnerable communities most require these services for situational awareness and information sharing, they might encounter even more significant disruptions than other groups . This paradox emphasizes the need for targeted interventions and strategies. Ensuring that these vulnerable communities have access to reliable internet services during disasters should be a priority.

These findings bring into focus a pressing need for policymakers, city planners, and internet service providers to re-evaluate infrastructure strategies. Enhancing the robustness of internet service in vulnerable areas, developing contingency plans for rapid response to outages, and perhaps even considering localized, community-based internet solutions could be potential steps forward.

## 4. Concluding Remarks

Evaluating the impact of hazards on infrastructure is critical for hazard monitoring and enhancing urban resilience. However, the disruption of information and communication technology (ICT) service—a crucial component of contemporary communication and urban operations—has long been overlooked by previous research. To address these gaps, this study innovatively utilized observational data on internet performance to delve into the impacts of hazard severity on internet disruption during two major US hazard events: Hurricane Harvey and Winter Storm Uri.

Our findings show that the hazard events caused significant disruptions of regional internet connectivity. Time series analysis indicates a strong correlation between the magnitude of disruption and hazard severity. Internet disruptions worsen with intensifying hazard severity and eventually return to baseline levels post-hazard. However, spatial analyses highlight that regions with pronounced hazard severity don't necessarily coincide with zones experiencing the highest internet disruptions. Areas not directly impacted by hazards can also experience severe disruption, demonstrating the repercussions of hazards extend beyond the immediate area of impact, affecting regions that might seem safe or untouched at first glance.

This temporal synchronization, juxtaposed with spatial discrepancies, underscores the complex relationships between hazard severity and internet disruption. Additionally, our sociodemographic inquiry sheds light on the disparate impacts of hazards across communities. The data reveals that the living conditions of the vulnerable communities may be further aggravated due to their already challenging situations, which call for special attention for prioritized strategies for disaster mitigation and interventions. Our research contributes to the body of knowledge in disaster prevention and urban resilience. It harnesses a quantitative observational dataset to elucidate the repercussions of hazards on the stability of ICT service. To the best of our knowledge, ours is the first study where the nexus between internet disruption and hazard severity is empirically and quantitatively examined. Our sociodemographic investigations further underscore the multifaceted relationship between vulnerable communities and the extent of internet disruption. The findings confirm the prevalent notion that vulnerable groups endure the worst of internet disruptions during significant events, offering insights into environmental inequities in disaster contexts. Practically, our discoveries provide invaluable perspectives on the disruptions, connectivity lapses, and recovery trajectories of internet performance in relation to hazard severity, which are pivotal for urban planners and infrastructure operators, aiding in the formulation of timely and effective strategies to mitigate hazard impacts and strengthen urban resilience.

The findings of this research point to several promising avenues for future exploration. While the current study was confined to two events pertaining to hurricanes and winter storms, restricted by data availability, future endeavors could seek to broaden the dataset. This expanded study would encompass a more varied range of hazard events, offering a comprehensive view of internet disruptions across different disaster contexts. Furthermore, deeper analysis should be undertaken to uncover the regional determinants that account for the spatial variances in internet resilience. Such insights would be instrumental in understanding why certain regions exhibit greater resilience compared to others, with the objective of identifying protective or mitigating factors.

**Acknowledgements**

We would like to acknowledge the data support from Ookla. This material is based in part upon work supported by the National Science Foundation under Grant CMMI-1846069 (CAREER). Any opinions, findings, conclusions, or recommendations expressed in this material are those of the authors and do not necessarily reflect the views of the National Science Foundation, or Ookla.